\documentclass[a4paper,11pt]{article}
\pdfoutput=1 

\usepackage{jinstpub} 
                     
\usepackage{siunitx}
\usepackage{multirow}
\newcommand{\neqcm}{\ensuremath{\mathrm{n}_{\mathrm{eq}}/\mathrm{cm}^2}}

\title{Performance of Irradiated RD53A 3D Pixel Sensors}

\author[a,1]{S.~Terzo,\note{Corresponding author.}}
\author[a]{M.~Chmeissani,}
\author[a]{G.~Giannini,}
\author[a,b]{S.~Grinstein,}
\author[c]{M.~Manna,}
\author[c]{G.~Pellegrini,}
\author[c]{D.~Quirion,}
\author[a]{D.~Vazquez~Furelos}

\affiliation[a]{Institut de F\'isica d'Altes Energies (IFAE), The Barcelona Institute of Science and Technology, Edifici CN, UAB campus, 08193 Bellaterra (Barcelona), Spain}
\affiliation[b]{Instituci\'o Catalana de Recerca i Estudis Avan\c{c}ats (ICREA), \\
Pg. Llu\'{i}s Companys 23, 08010 Barcelona, Spain}
\affiliation[c]{Centro Nacional de Microelectronica (CNM-IMB-CSIC), \\
08193 Bellaterra (Barcelona), Spain }

\emailAdd{sterzo@ifae.es}

\abstract{
The ATLAS experiment at the LHC will replace its current inner
tracker system for the HL-LHC era. 3D silicon pixel sensors are
being considered as radiation-hard candidates for the innermost layers
of the new fully silicon-based tracking detector. 
3D sensors with a small pixel size of \SI{50x50}{\um} and \SI{25x100}{\um}
compatible with the first prototype ASIC for the HL-LHC, the RD53A chip, 
have been studied in beam tests after uniform irradiation to \SI{5e15}{\neqcm{}}.
An operation voltage of only \SI{50}{V} is needed to achieve a \SI{97}{\%} hit efficiency after this fluence.
}

\keywords{Solid state detectors, Radiation-hard detectors, Particle tracking detectors} 

\begin{document}
\maketitle
\flushbottom

\section{Introduction}
\label{sec:intro}

In order to test the predictions of the Standard Model of particle physics, the Large Hadron Collider (LHC) experiments need to identify and determine the path and origin of the particles that are produced in the LHC proton-proton collisions. 
Silicon pixel detectors are especially important for the
precise determination of tracks and vertices, enabling the selection 
of interesting events through the identification of b-jets (b-tagging). 
To further probe the energy frontier after 2025, the High Luminosity LHC 
(HL-LHC) project aims at improving the accelerator performance. At the 
same time, the pixel detectors of the LHC experiments have to be upgraded 
to maintain their potential for discoveries.

The innermost layers of the tracking detectors at the HL-LHC 
experiments will have 
to sustain unprecedented radiation levels with \SI{1}{MeV} neutron 
equivalent fluences up to \SIrange{1}{2e16}{\neqcm{}} and an 
increased particle rate. 
The 3D pixel sensor technology has been selected as the baseline option for 
the innermost layer of the ATLAS pixel tracker (ITk)~\cite{ATLAS_Pixel_TDR} 
due to its excellent results~\cite{Lange1.4E16}, in terms of radiation 
hardness and power dissipation. 
In 3D pixel sensors, n- and p-type column-like electrodes penetrate the silicon substrate 
perpendicularly to the surface defining the pixel configuration. This decouples the
device active thickness from the electrode separation, which can be chosen to be significantly 
smaller than for the standard planar sensors, thus reducing trapping effects after irradiation. 

Previous studies of 3D sensor hybrid devices with small
pixel cells of \SI{50x50}{\um} and \SI{25x100}{\um}, irradiated to high fluences were conducted using the ATLAS FE-I4 readout chip~\cite{fei4}. These devices belonged to the first generation of 3D sensors for high energy physics experiments (successfully used in the ATLAS IBL and AFP sub-detector systems), which 
were produced in a \SI{230}{\um} thick active bulk and with the columns etched from both sides of the substrate.
Since then, the RD53A chip~\cite{rd53a}, the first ASIC prototype developed for the pixel detectors of the HL-LHC experiments, became available. Correspondingly, the second generation 3D sensors were produced. These sensors have an active thickness of \SI{150}{\um} and the columns are etched from the same side of the substrate.
This paper presents the first results of hit reconstruction efficiency 
and power dissipation performance of 3D sensors read out with the RD53A ASIC prototype after uniform irradiation to \SI{5e15}{\neqcm{}}. 
 
\section{The RD53A 3D devices under test}
The RD53A readout chip presents a matrix of \SI{400x192}{\um} pixels of 
\SI{50x50}{\um} with a total size of \SI{20.0x11.8}{mm}. The ASIC, 
fabricated in the \SI{65}{\nm} CMOS technology, includes three different 
front-end designs to allow performance comparisons. 
These are the differential, linear and synchronous front-ends. 
The results presented in this paper were obtained using the 
differential and the linear front-ends.

\begin{figure}[htbp]
\centering
\includegraphics[width=0.45\textwidth]{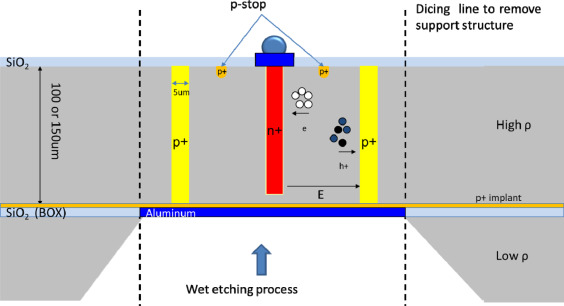}
\caption{\label{fig:soi} Schematic representation of the CNM single-side process on SOI~\cite{cnm-SoI-rd53a}.}
\end{figure}

The 3D sensors were produced at CNM~\cite{cnm-SoI-rd53a} on \SI{100}{\mm} SOI wafers which had a total thickness of \SI{450}{\um}. These are part 
of a dedicated R\&D run (9761) for the ATLAS ITk Pixel detector. A schematic representation of the 3D sensor technology of this CNM run is shown in figure~\ref{fig:soi}.
The \SI{150}{\um} active thickness of the sensors is defined by a p-type backside implant and a buried silicon oxide layer that separates the high resistivity (between \SIlist{1000;5000}{\Omega\cm}) p-type bulk, from the low resistivity substrate that acts as a handle wafer to keep the structure rigidity during the fabrication cycle.
The thin layer of silicon oxide is used as an etch stop during wet etching process that removes the handle wafer. 
The columns that define the pixel cells were etched from the
same side (single-side process). The p-type columns have a target
depth of \SI{150}{\um} while the n-type columns are shorter to avoid reaching the p-type backside layer. 
The etchings of the p- and n-columns are critical processes which need micrometer accuracy to reach the backside implantation (p-type) or to remain at the right distance from it (n-type). Etching of p-type columns may use the buried oxide layer as an etch-stop, with a controlled over-etch.
The diameter of the columns is \SI{8}{\um} with p-stop 
isolation implants surrounding the n-type columns. 
The bias voltage is applied from the backside after the
low resistivity substrate is etched and an aluminum 
layer is deposited to contact the p-type backside implant.

\begin{figure}[htbp]
\centering
\includegraphics[width=0.9\textwidth]{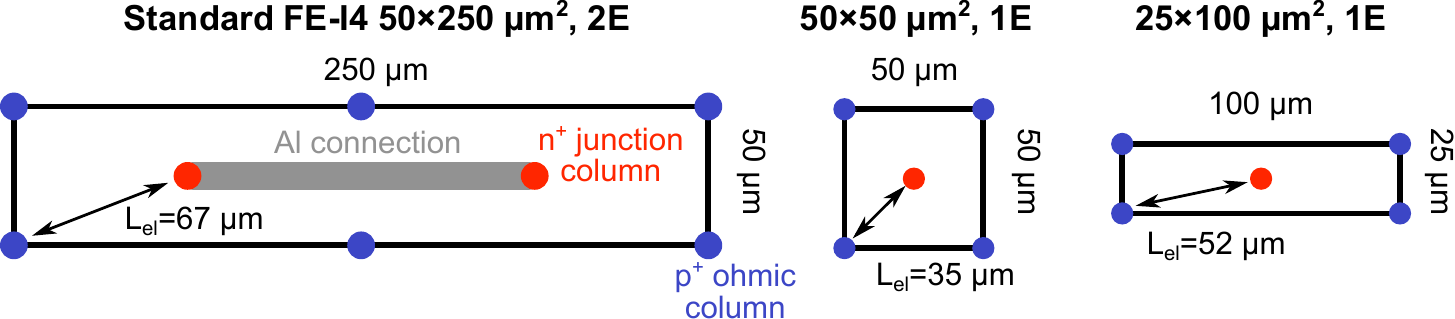}
\caption{\label{fig:smallpitch} Schematic layout of the different 3D pixel cell geometries. From left to right: a standard IBL/AFP FE-I4 pixel cell with \SI{50x250}{\um} and two electrode configuration (2E), a \SI{50x50}{\um} and a \SI{25x100}{\um} RD53A pixel cells both with one electrode configuration (1E). The n+ readout columns are shown in red and the p-columns in blue.}
\end{figure}

Results from devices with pixel geometries of \SI{50x50}{\um} and \SI{25x100}{\um} are included in this paper.
Both geometries present one n-type electrode in the center and four 
p-type columns in the corners of each pixel cell (as shown in figure~\ref{fig:smallpitch}) and are compatible with the RD53A ASIC.

The hybridization process was carried out by IFAE~\cite{AFP-flip-chip}. 
The RD53A ASICs had tin-silver bumps deposited at IZM~\cite{IZM-germany}.
The under-bump metalization (UBM) of the 3D sensors was performed at CNM
using an electroless nickel-gold process~\cite{cnm-ubm}. 
Due to the small diameter opening of the pads of the 3D sensors (\SI{12}{\um}), this method of UBM is not optimal and leads to areas of poor metal deposition. 
However, the method is fast and relatively inexpensive
compared to the alternative approach of electroplating. Since the target 
of this study is understanding the performance of the sensor, it is not 
critical to have all the channels of the ASIC connected.

The next key step of the hybridization process is flip-chipping, 
in which the sensor and the ASIC are thermo-compressed to connect
the readout channels and the sensor pixels. 
To improve connectivity the devices were reflowed with formic
acid. To verify that at least a large fraction of the channels were 
connected, the modules were inspected with X-rays. Though areas of 
disconnected bumps were observed, enough connected pixels to be able 
to carry out the desired performance studies were identified.


After flip-chipping the modules are assembled on PCBs designed by the University of Bonn for the characterisation.

\section{Device irradiation and beam tests}

Two assembled modules with different sensor geometries of \SI{50x50}{\um} and \SI{25x100}{\um} were irradiated at the cyclotron of the Karlsruher Institut f\"ur Technologie (KIT)~\cite{kitirr} with \SI{23}{MeV} protons to a fluence of about \SI{5e15}{\neqcm{}}, equivalent to a Total Ionisation Dose (TID) of about \SI{750}{Mrad}. The irradiation was performed at a controlled temperature of \SI{-30}{\celsius} and without powering the chip. The uncertainty on the measurement of the final fluence is of \SI{10}{\%}. 

\begin{figure}[htbp]
\centering
\includegraphics[width=.47\textwidth]{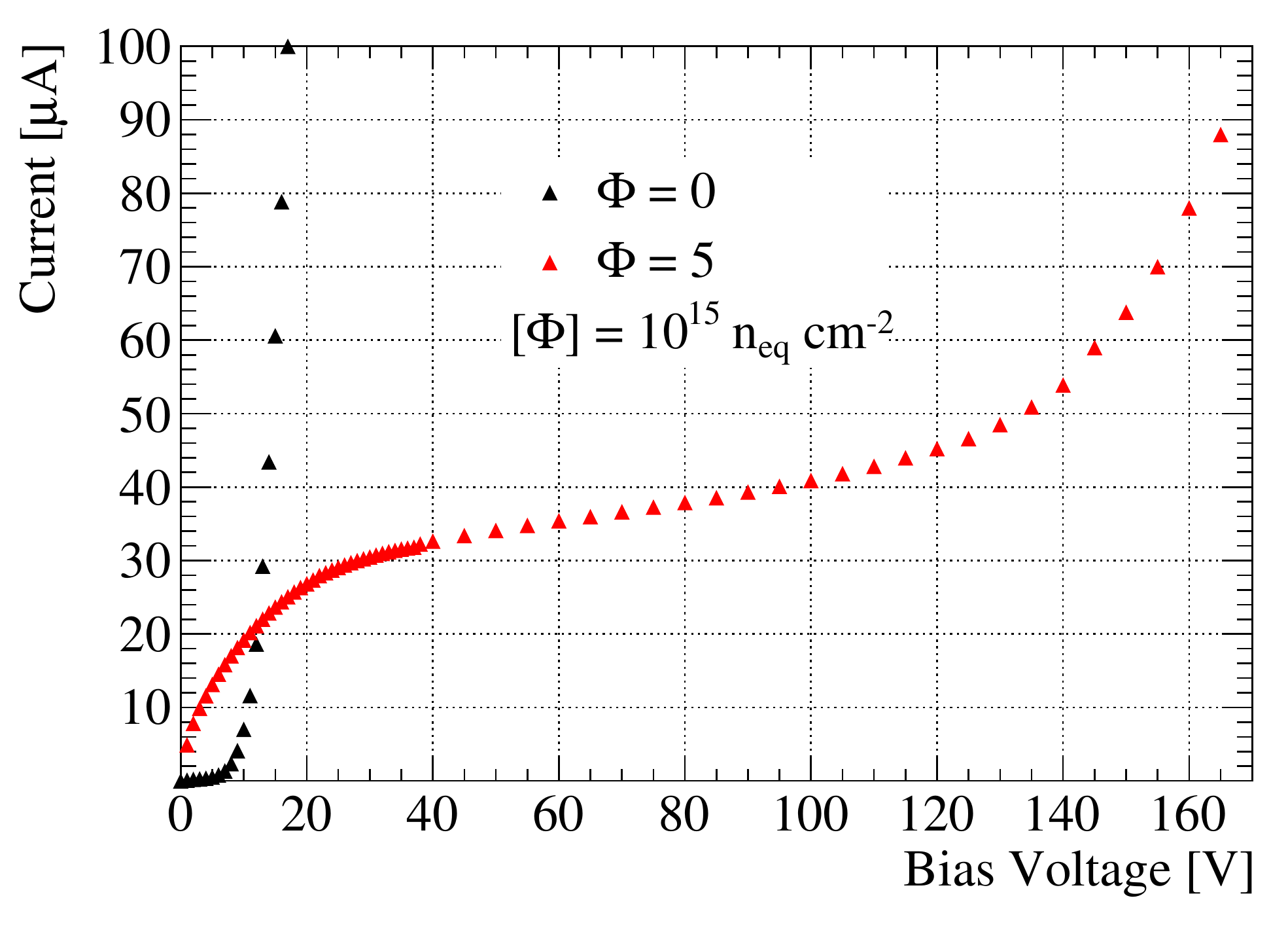}
\qquad
\includegraphics[width=.47\textwidth,origin=c]{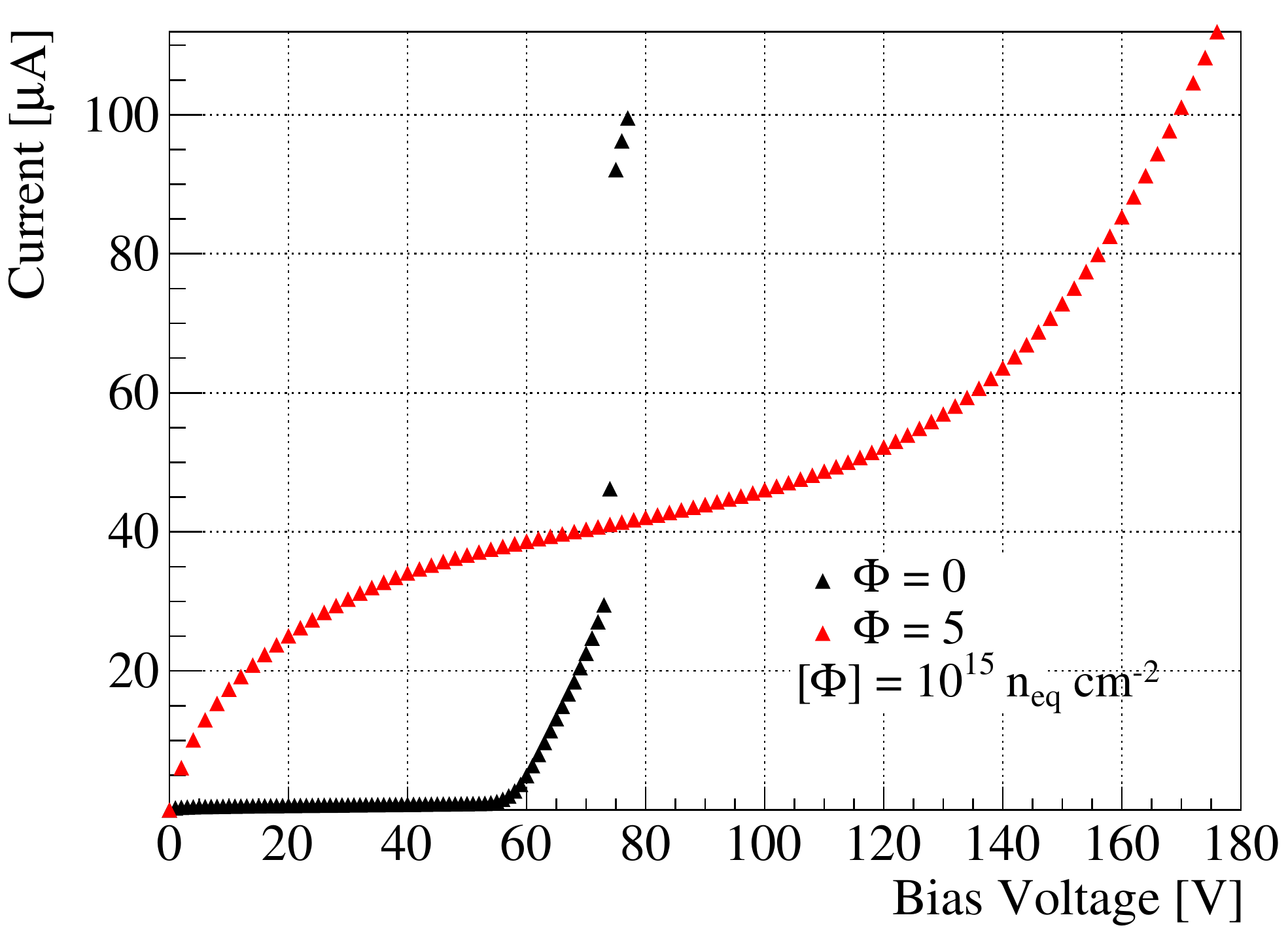}
\caption{\label{fig:iv} Leakage current as a function of the bias voltage for the \SI{50x50}{\um} sensor (left) and the \SI{25x100}{\um} sensor (right). The measurements were performed before irradiation (black) and after irradiation to \SI{5e15}{\neqcm{}} (red) at a stable temperature of \SI{20}{\celsius} and \SI{-25}{\celsius}, respectively.}
\end{figure}

The leakage current of the two sensors as a function of the bias voltage is compared before and after irradiation in figure~\ref{fig:iv}. Before irradiation the \SI{50x50}{\um} sensor had a very early breakdown with a steep rise of the current already between \SIlist{5;10}{V} while the \SI{25x100}{\um} sensor presented a breakdown around \SI{60}{V}. Nevertheless, after irradiation the two sensors show a similar electrical behaviour with a smooth breakdown around \SI{140}{V} and a leakage current between \SIlist{30;60}{\uA} in the range from about \SI{35}{V} to the breakdown.

Beam test measurements of the two modules before and after irradiation were performed at the CERN SPS beam line H6 with \SI{120}{GeV} pions. For particle tracking a EUDET-type telescope made of six MIMOSA26 tracking planes was used~\cite{eudet}. An ATLAS FE-I4 pixel module was employed as time reference to select tracks within \SI{25}{ns} from the trigger signal given by a set of scintillators. The RD53A chips were tuned and operated at beam test with the BDAQ53 readout system~\cite{bdaq53} developed by the University of Bonn. 

Both the linear and the differential front-ends were measured for the two modules. Nevertheless, due to a significant disconnected area at the right edge of the \SI{25x100}{\um} module, the results on the differential front-end are limited to a small fraction of the total pixels.

\section{Beam test results}
\begin{figure}[htbp]
\centering
\includegraphics[width=.48\textwidth]{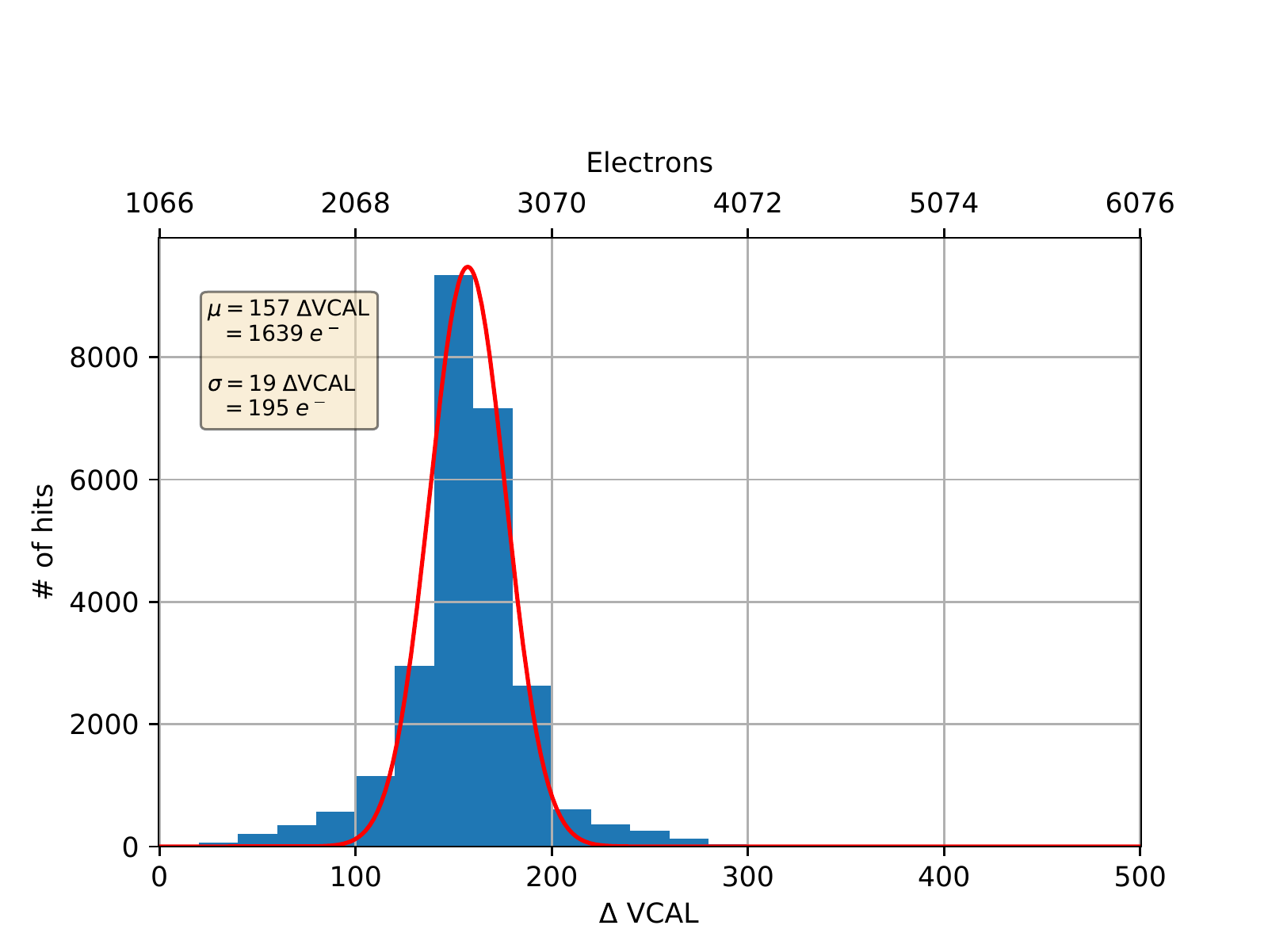}
\qquad
\includegraphics[width=.44\textwidth,origin=c]{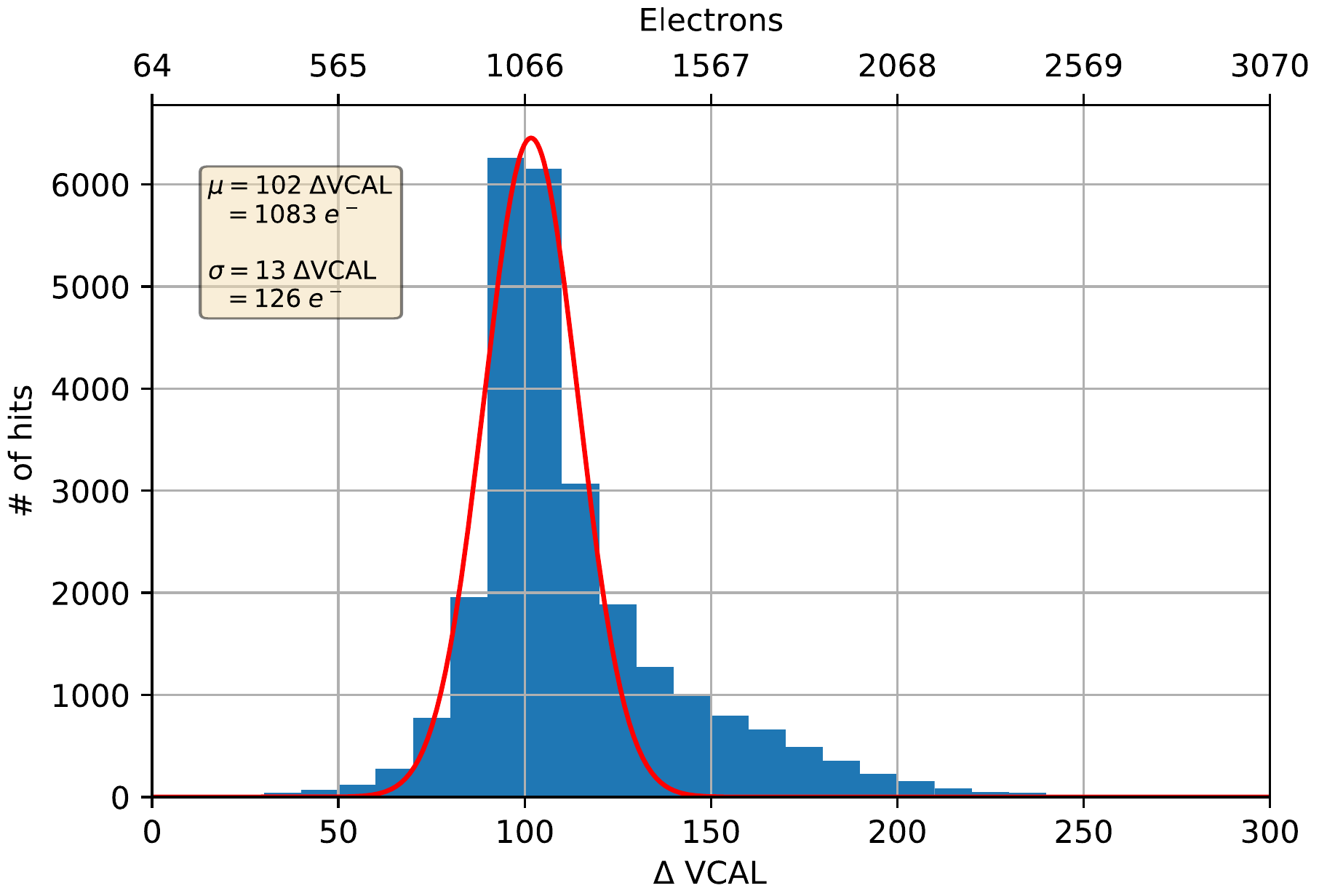}
\caption{\label{fig:thrdist} Threshold distributions of the Linear front-end of the \SI{50x50}{\um} (left) and the \SI{25x100}{\um} RD53A modules (right) after uniform irradiation at KIT to a particle fluence of \SI{5e15}{\neqcm{}}.}
\end{figure}

The two modules were tuned aiming at a threshold of \SI{1}{ke}, nevertheless, to reduce the number of noisy pixels especially after irradiation the resulting average thresholds at which the modules were operated are between \SIlist{1;1.6}{ke}. Table~\ref{tab:tune} reports the mean value of the Gaussian fit over the threshold distribution of all pixels in each of the two front-ends. It has to be noted that as a consequence of the large TID received by the chips in the KIT irradiation, which is beyond the RD53A specifications of \SI{500}{Mrad}, the threshold distribution presented large tails as shown in figure~\ref{fig:thrdist}.

\begin{table}[htbp]
\centering
\caption{\label{tab:tune} Tuning before and after irradiation of the RD53A modules during particle beam tests.}
\smallskip
\begin{tabular}{|c|c|c|c|c|}
\hline
Sensor pitch & Fluence              & \multirow{2}{*}{Front End} & \multicolumn{2}{c|}{Threshold}\\
\cline{4-5}
~[\si{\um^2}]~ & [\SI{e15}{\neqcm{}}] &           & Mean [e] & Sigma [e]\\
\hline
\multirow{4}{*}{\SI{50x50}{}} & \multirow{2}{*}{0} & Linear & 1111 & 101\\
                              &                    & Differential & 1066 & 51\\
                              \cline{2-5}
                              & \multirow{2}{*}{5} & Linear & 1639 & 195\\
                              &                              & Differential & 1060 & 34 \\ 
  \hline
\multirow{4}{*}{\SI{25x100}{}} & \multirow{2}{*}{0} & Linear & 1075 & 83\\
 &  & Differential & 1127 & 66\\
 \cline{2-5}
                               & \multirow{2}{*}{5} & Linear & 1083 & 126\\
                                &  & Differential & 1313 & 147 \\
\hline
\end{tabular}
\end{table}

\begin{figure}[htbp]
\centering
\includegraphics[width=.47\textwidth]{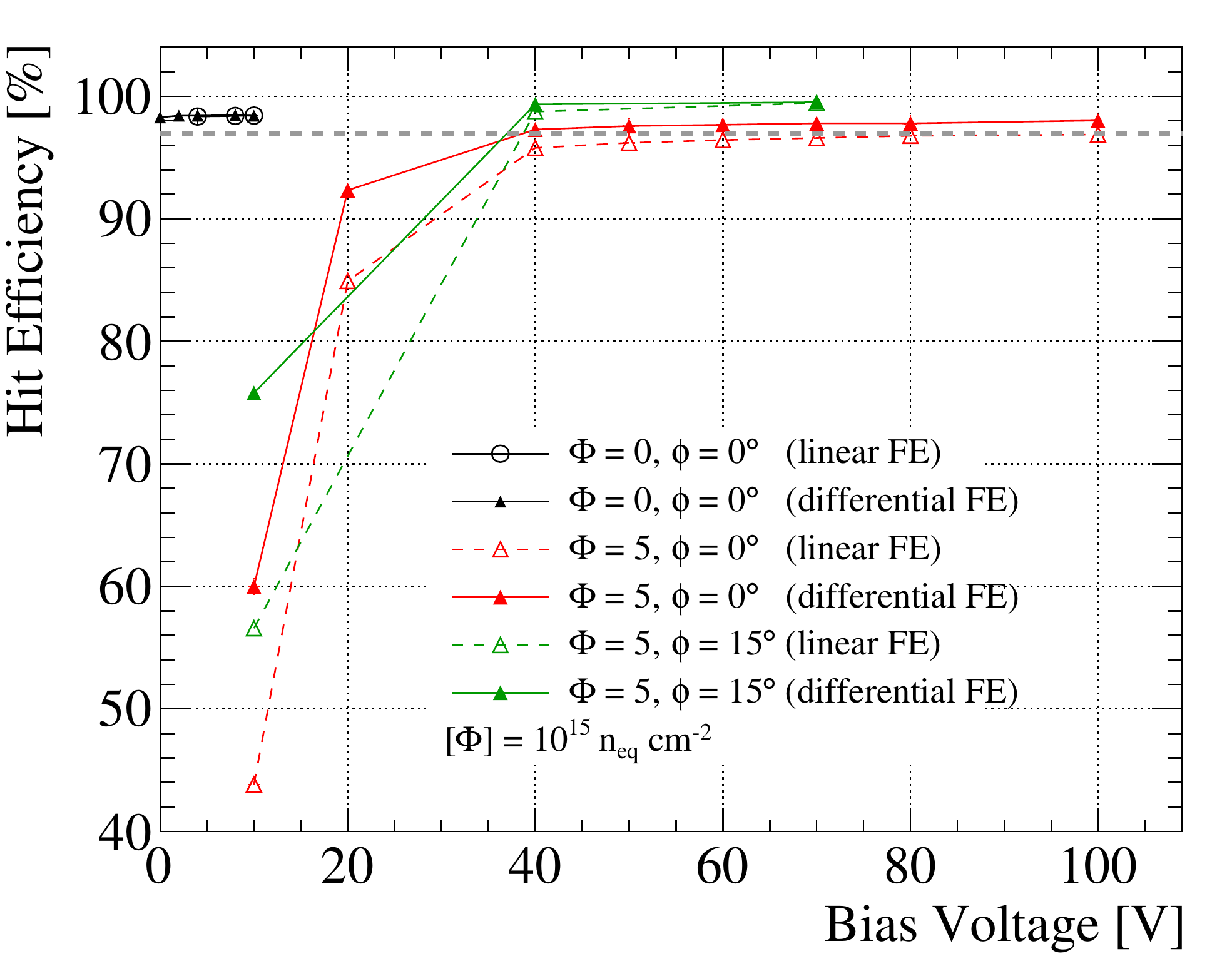}
\qquad
\includegraphics[width=.47\textwidth,origin=c]{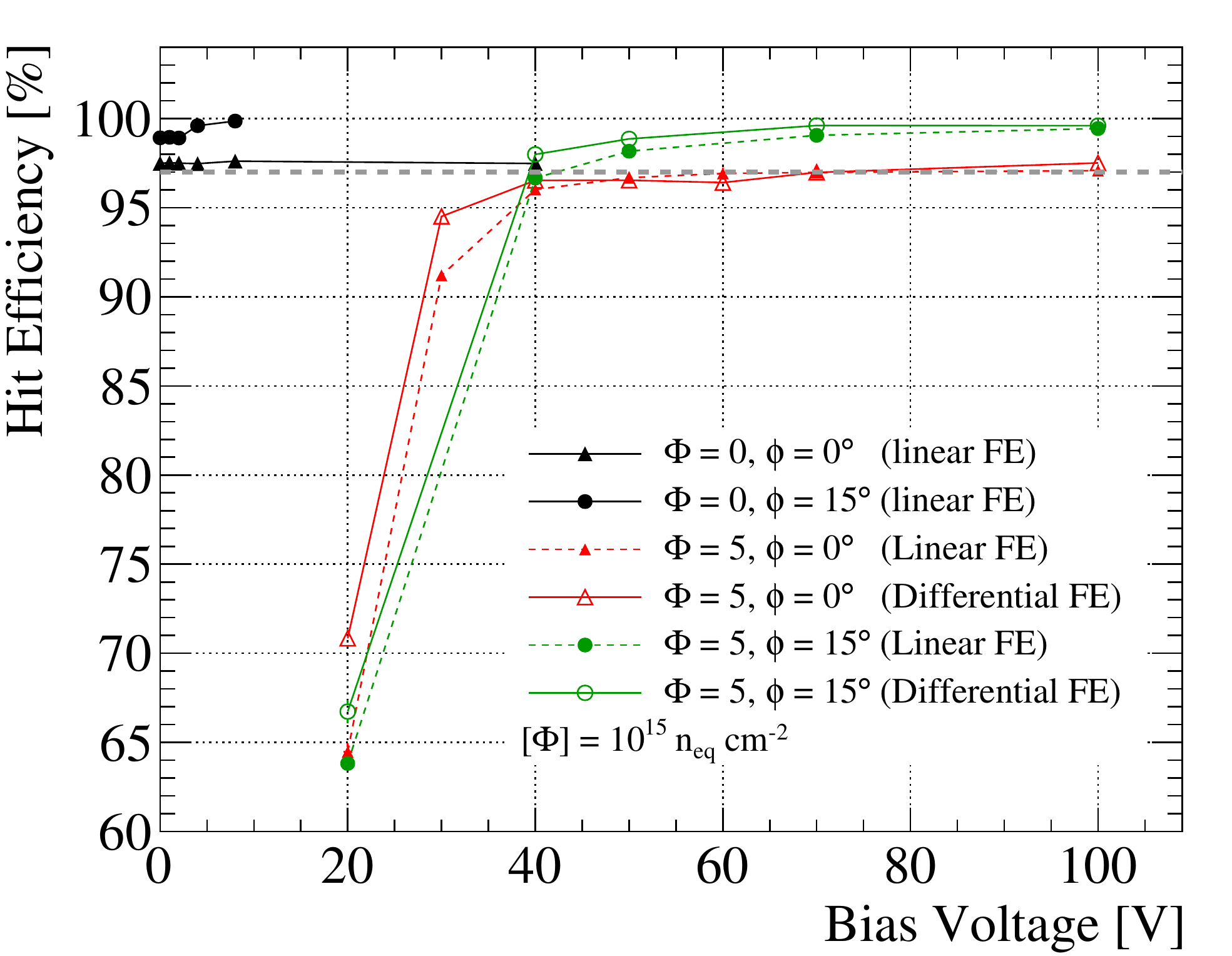}
\caption{\label{fig:effvsbias} Hit efficiency as a function of the bias voltage for the \SI{50x50}{\um} (left) and the \SI{25x100}{\um} modules (right). The dashed line indicates the ATLAS benchmark hit efficiency of \SI{97}{\%} for the ATLAS pixel detector. A systematic uncertainty of \SI{0.3}{\%} is associated to all measurements.}
\end{figure}

Results of the hit efficiency measured before and after irradiation as a function of the bias voltage are shown for the two measured modules in figure~\ref{fig:effvsbias}. Before irradiation the hit efficiency is larger than \SI{97}{\%} even without bias voltage applied and saturates around \SI{98}{\%}. After irradiation a hit efficiency higher than \SI{96}{\%} is reached for both sensor geometries at \SIlist{40}{V} and it stabilises  around \SI{97}{\%} up to the maximum measured bias of \SI{100}{V}. 

\begin{figure}[htbp]
\centering
\includegraphics[width=.47\textwidth]{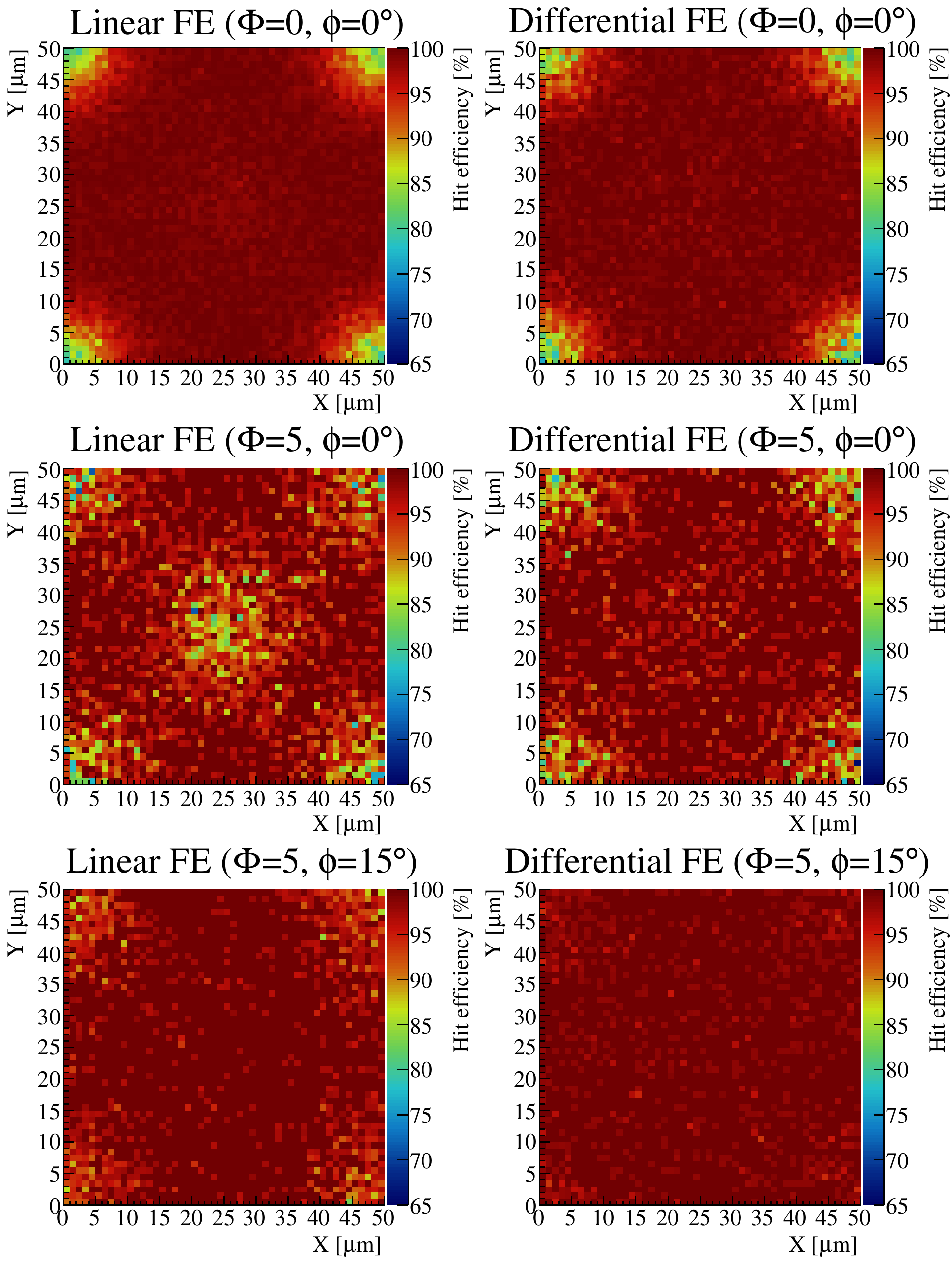}
\qquad
\includegraphics[width=.47\textwidth,origin=c]{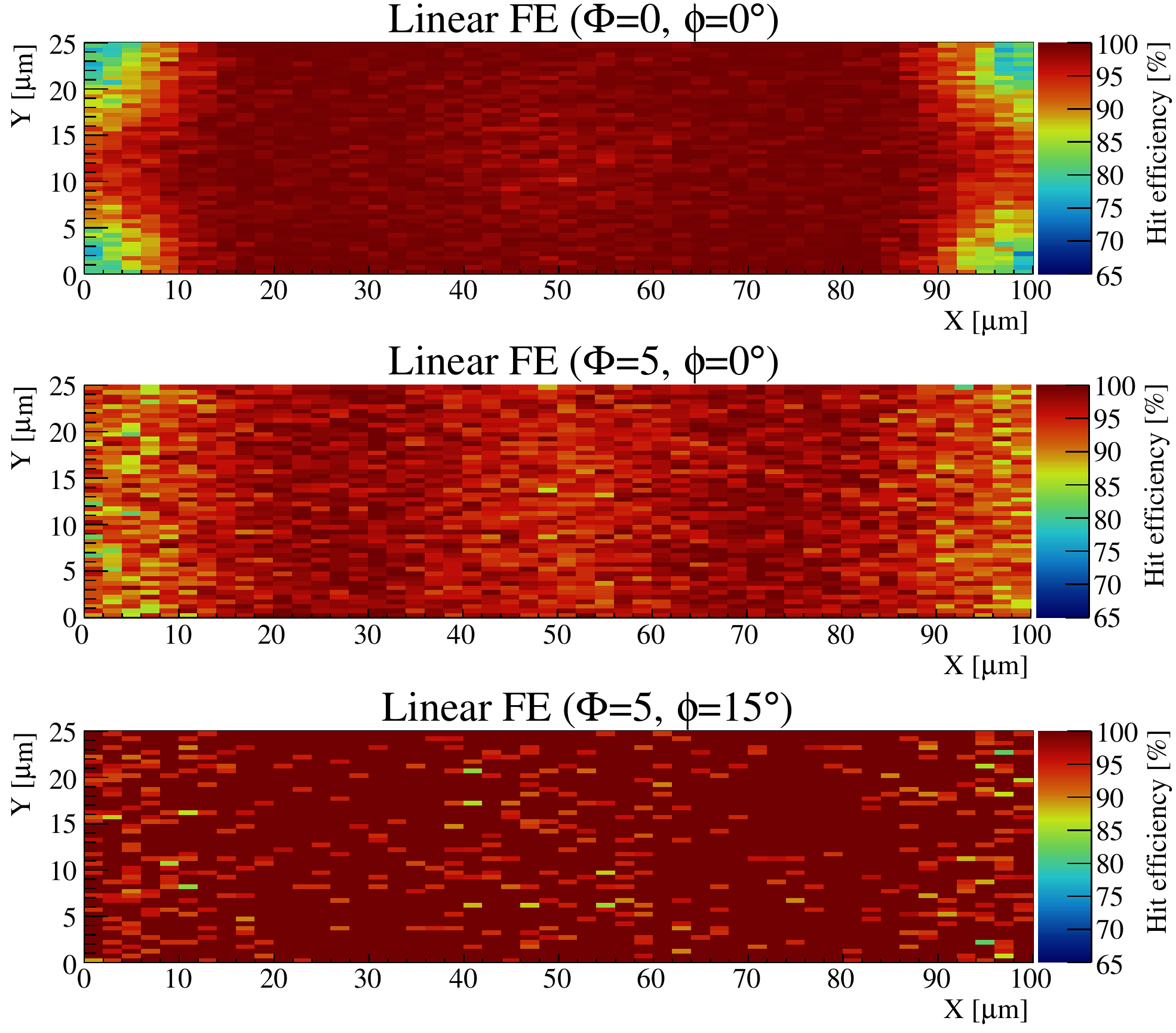}
\caption{\label{fig:effmaps} Hit efficiency over one pixel surface for \SI{50x50}{\um} (Linear front-end on the left, Differential front-end in the centre) and the \SI{25x100}{\um} (Linear front-end on the right) modules. From top to bottom the maps show modules measured before irradiation at perpendicular beam incidence ($\mathrm{\Phi=0}$, $\mathrm{\phi=0}$), after irradiation at perpendicular beam incidence ($\mathrm{\Phi=5}$, $\mathrm{\phi=0}$) and after irradiation with \SI{15}{\degree} beam incidence angle ($\mathrm{\Phi=5}$, $\mathrm{\phi=15}$). The hit efficiency maps of the pixel cells are obtained displaying the reconstructed track impact point expressed in pixel coordinates and projecting the data for all identical structures onto the same image.}
\end{figure}

Both before and after irradiation the efficiency lost is mostly due to the passing through p-columns in the corners of the pixel cell which adds up to the charge sharing between neighbouring pixels as can be seen in the maps of figure~\ref{fig:effmaps}. On the other hand the n-column in the centre of the pixel cells is not fully penetrating the active bulk, but leaves about \SI{20}{\um} of active silicon in which enough charge is generated to detect a particle. No significant inefficiency is thus observed in the central region before irradiation, but only after irradiation and especially as the threshold approaches values close to about \SI{1.6}{ke} i.e. the expected Most Probable Value of the charge distribution generated in \SI{20}{\um} of silicon by a minimum ionising particle. 

By tilting the modules \SI{15}{\degree} the hit efficiency increases with respect to the perpendicular case and saturates over \SI{99}{\%} both before and after irradiation. This is due to the fact that particles are not passing anymore completely through the inactive columns. The effect can be observed in the bottom plots of figure~\ref{fig:effmaps} where the hit efficiency over the pixel surface for tilted devices becomes almost uniform. A tilted configuration is also foreseen for the pixel modules in the final ITk detector layout~\cite{ATLAS_Pixel_TDR}, thus the effect of the passing columns of 3D sensors for the detection performance will be negligible. 

The \SI{50x50}{\um} module after irradiation presents a difference in the efficiency between linear and differential front-ends 
that is consistent with the different thresholds obtained in the tuning of these two front-ends. 
In the \SI{25x100}{\um} instead, the linear front-end performs slightly worse than the differential one despite the lower average threshold. This is given by the fact that the few connected pixels in the differential front-end mostly belong to the core of the threshold distribution and the effect of the large tails is thus not accounted for.

\begin{figure}[htbp]
\centering
\includegraphics[width=.55\textwidth]{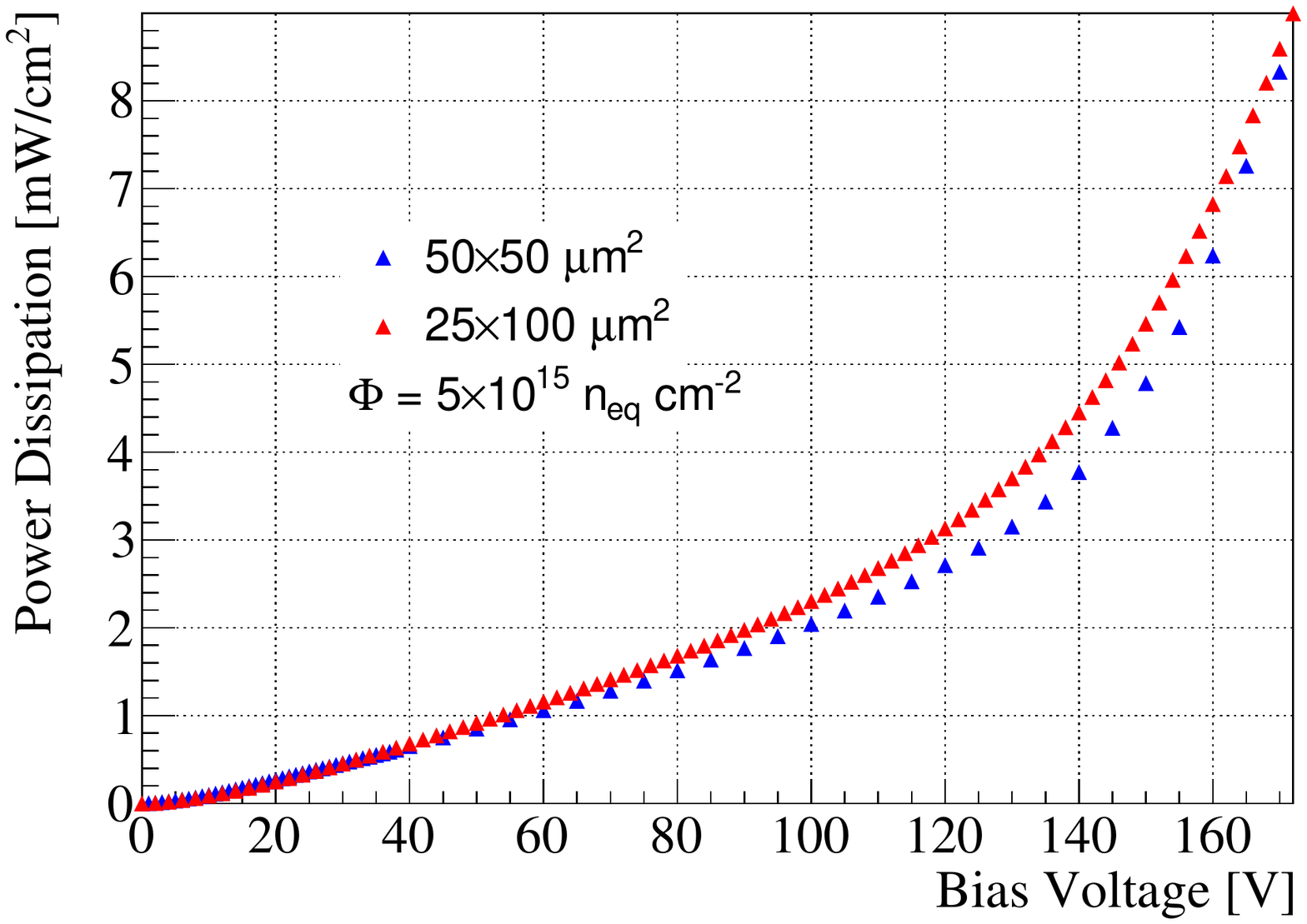}
\caption{\label{fig:powdiss} Power dissipation as a function of the applied bias voltage for \SI{50x50}{\um} (blue) and \SI{25x100}{\um} (red)  sensors after uniform irradiation.}
\end{figure}

Since the irradiation at KIT was uniform over the sensor surface, it is possible to estimate the power dissipation of the two measured modules at the operational voltage i.e. the voltage at which the hit efficiency reaches \SI{97}{\%}. This can be calculated from the power per area as function of the voltage shown in figure~\ref{fig:powdiss}. Considering an operational voltage of \SIrange{40}{50}{V} for both pixel geometries a very low power dissipation of less than \SI{1}{mW/cm^{2}} at \SI{-25}{\celsius} is measured. These results are consistent with previous measurements of thicker small pitch 3D sensor prototypes~\cite{Lange1.4E16}. Thus the sensor thickness introduces a negligible effect on the power dissipation performance of the devices at these radiation levels.

\section{Conclusions and outlook}
Novel radiation-hard small-pitch 3D pixel sensors compatible with the new RD53A chip and featuring a \SI{150}{\um} thick active substrate have been produced at CNM for the upgrade of the ATLAS ITk for the HL-LHC era. RD53A pixel modules employing 3D sensors with \SI{50x50}{\um} and \SI{25x100}{\um} 1E pixel cells have been characterised at beam tests before and, for the first time, after uniform irradiation to a particle fluence of \SI{5e15}{\neqcm{}}.

Before irradiation both sensor geometries have shown a hit efficiency higher than \SI{97}{\%} at perpendicular beam incidence even without any bias voltage applied. After irradiation the measured hit efficiency at perpendicular beam incidence is higher than \SI{96}{\%} when applying a bias of \SI{40}{V} or more. By tilting the devices \SI{15}{\degree} with respect to the beam axis the hit efficiency reaches more than \SI{99}{\%} both before and after irradiation. The power dissipation of irradiated 3D sensors measured at \SI{-25}{\degree} with an operation voltage of \SI{50}{V} is less than \SI{1}{mW/cm^{2}}.

Further irradiation of RD53A modules assembled with 3D sensors from the CNM run (9761) are foreseen up to the HL-LHC total flunces expected for the innermost layer of ITk (of the order of \SI{e16}{\neqcm{}}).
Moreover, a new 3D sensor production has been recently completed at CNM using Silicon on Silicon (Si-Si) wafers with \SI{150}{\um} active thickness. This also includes \SI{50x50}{\um} and \SI{25x100}{\um} pixel sensor geometries compatible with RD53A chips. With respect to SOI, the Si-Si approach allows direct biasing from the conductive handle wafer, thus avoiding the need to etch the backside. In addition, since the silicon oxide layer between the active and the handle layers is not required, there is a greater flexibility in the choice of the DRIE receipt. Because of these advantages the Si-Si wafers will be the technology of choice for the final production of 3D sensors for ITk.

\acknowledgments
This work was partially funded by: MINECO, Spanish Government, under grants FPA2015-69260-C3-2-R, FPA2015-69260-C3-3-R and SEV-2016-0588 (Severo Ochoa excellence programme); and by the H2020 project AIDA-2020, GA no. 654168. The authors would like to thank A. Rummler, M. Bomben, J. Lange and the other ATLAS ITk beam test participants for great support and discussions at the beam test as well as A. Dierlamm and F. B{\"o}gelspacher for the excellent support with the irradiations at KIT.


\end{document}